\documentclass[aps,prb,twocolumn,showpacs,amsmath,amssymb,superscriptaddress]{revtex4-1}
\usepackage{graphicx}
\usepackage{bm}
\usepackage{epsfig}
\usepackage{epstopdf}	
\include{graphic}	

\begin{document}
\title{Travelling photons mediated interactions between a magnon mode and a cavity photon mode}

\author{J. W.~Rao}\email{jinweir@myumanitoba.ca;}
\affiliation{Department of Physics and Astronomy, University of Manitoba, Winnipeg, Canada R3T 2N2}
\affiliation{The Key Lab for Magnetism and Magnetic Materials of Ministry of Education, Lanzhou University, Lanzhou 730000, China}
\author{Y. P.~Wang}
\affiliation{Department of Physics and Astronomy, University of Manitoba, Winnipeg, Canada R3T 2N2}
\author{Y.~Yang}
\affiliation{Department of Physics and Astronomy, University of Manitoba, Winnipeg, Canada R3T 2N2}
\author{T.~Yu}
\affiliation{Kavli Institute of NanoScience, Delft University of Technology, 2628 CJ Delft, The Netherlands}
\author{Y. S.~Gui}
\affiliation{Department of Physics and Astronomy, University of Manitoba, Winnipeg, Canada R3T 2N2}
\author{X. L.~Fan}
\affiliation{The Key Lab for Magnetism and Magnetic Materials of Ministry of Education, Lanzhou University, Lanzhou 730000, China}
\author{D. S.~Xue}
\affiliation{The Key Lab for Magnetism and Magnetic Materials of Ministry of Education, Lanzhou University, Lanzhou 730000, China}
\author{C.-M.~Hu}\email{Can-Ming.Hu@umanitoba.ca;}
\affiliation{Department of Physics and Astronomy, University of Manitoba, Winnipeg, Canada R3T 2N2}

\begin{abstract}

We systematically study the indirect interaction between a magnon mode and a cavity photon mode mediated by travelling photons of a waveguide. From a general Hamiltonian, we derive the effective coupling strength between two separated modes, and obtain the theoretical expression of system's transmission. Accordingly, we design an experimental set-up consisting of a shield cavity photon mode, microstrip line and a magnon system to test our theoretical predictions. From measured transmission spectra, indirect interaction, as well as mode hybridization, between two modes can be observed. All experimental observations support our theoretical predictions. In this work, we clarify the mechanism of travelling photon mediated interactions between two separate modes. Even without spatial mode overlap, two separated modes can still couple with each other through their correlated dissipations into a mutual travelling photon bus. This conclusion may help us understand the recently discovered dissipative coupling effect in cavity magnonics systems. Additionally, the physics and technique developed in this work may benefit us in designing new hybrid systems based on the waveguide magnonics. 

\end{abstract}

\maketitle

\section{introduction}

Engineering hybrid systems which combine complementary physical components is one of the central goals in quantum technology.\cite{Khitrova1999, Raimond2001, Xiang2013, Aspelmeyer2014, Kurizki2015} Following this approach in the past few years, the field of cavity magnonics has been developed through the coupling of collective spin excitations, i.e., magnons, with cavity photons.\cite{Soykal2010, Dany2019} This system was first proposed by Soykal and Flatt\'{e} in 2010 \cite{Soykal2010}, but only realized experimentally since 2013\cite{Huebl2013, Huebl2013, Zhang2014, Tabuchi2014, Goryachev2014, Lambert2015, Bai2015, Li2019, Hou2019}. So far, many technologies have been developed based on this versatile system, for example the gradient memory architecture\citep{Zhang2015}, single magnon detection\cite{Dany2017}, non-local spin current manipulation\cite{Bai2017}, cavity magnon polariton logic gate\cite{JW2019}, giant non-reciprocity\cite{Wang2019} etc.. In all these previous works, spatial mode overlap between the cavity microwave field and the magnon mode is essential to achieve strong photon-magnon interaction, and hence magnon systems are deliberately placed inside cavities. An inevitable problem is that all operations in cavity magnonics systems must take in account both the geometry and mode distribution of cavities. To get ride of these spatial limitations, coupling magnon system to other resonant components, including cavity photon, qubit and atom etc., through travelling photons may be a feasible solution.

In fact, this kind of indirect interactions mediated by travelling photons is a main object of research in waveguide quantum electrodynamics (QED). It has been verified in various systems including quantum dots\cite{Chang2006,Tudela2011}, atoms\cite{Vetsch2010,Goban2012,Kockum2018} and superconducting circuits\cite{Loo2013,Lalumiere2013,Mirhosseini2018}, but has not yet been tested in magnon system. In this work, we firstly extend the idea of waveguide QED into magnon system, and systematically study the interaction between a magnon mode and a separated cavity photon mode. At beginning, we construct a general Hamiltonian, from which theoretical explanation of the indirect interaction between two separated modes is given. This effect is sustained by travelling photons, and its intensity equals to the square root of the product of two modes' phase correlated dissipations. Accordingly, we design an experimental set-up consisting of a magnon system, microstripline and a shield cavity photon mode. From measured transmission spectra, mode hybridization between the magnon mode and the separated cavity photon mode can be observed, which exhibits a strong dependency in the separation of two modes. All these experimental observations well confirm our theoretical predictions. Our work clarifies the mechanism of dissipative coupling mediated by travelling photons, which is useful for us to understand the recently discovered dissipative coupling\cite{Wang2019,Harder2018,Grigoryan2018,Metelmann2015,Yang2019,Yu2019,Rao2019,Bhoi2019,Boventer2019,Yao2019} and giant non-reciprocity\cite{Wang2019} in cavity magnonics systems. Additionally, the physics and technique developed in our work may help us design new hybrid system based on waveguide magnonics. 

\section{Theoretical model}

In cavity magnonics systems, strong interactions between cavity photon and magnon arises from the spatial overlap of the cavity microwave field and the magnon mode. However, in a waveguide magnonics system, such a direct overlap is unnecessary, because a magnon mode in waveguide can be coupled to other components, including a separated cavity photon mode, through their correlated dissipations into a mutual waveguide photon bus. 

\subsection{A general Hamiltonian for the waveguide magnonics}

To illustrate this effect, we construct a general Hamiltonian in the form of:
\begin{eqnarray}
&H&=\hbar\omega_m\hat{m}^\dag\hat{m}+\hbar\omega_a\hat{a}^\dag\hat{a}+\int\hbar\omega_k\hat{p_k}^\dag\hat{p_k}dk+\int \hbar[\lambda_me^{i\phi} \nonumber \\
&\quad&(\hat{m}+\hat{m}^\dag)(\hat{p_k}+\hat{p_k}^\dag)+\lambda_a(\hat{a}+\hat{a}^\dag)(\hat{p_k}+\hat{p_k}^\dag)]dk \label{Eq1}
\end{eqnarray}
where $\hat{m}(\hat{m}^\dag)$ and $\hat{a}(\hat{a}^\dag)$ are the annihilation (creation) operators of the magnon mode and the other resonant mode, respectively. $\omega_m$ and $\omega_a$ correspond to their uncoupled mode frequencies. Considering these two mode are separated with each other in a waveguide, there is no direct mode overlap between them. Therefore, the direct coupling effect between them has not been considered in this Hamiltonian. The third term of Hamiltonian represents travelling photons in a waveguide, which is an integral of wave vector over the whole real domain ($-\infty$ to $\infty$). $\hat{p_k}$ is the boson annihilation operator of the travelling photon with $[\hat{p_k},\hat{p}_{k'}]=\delta(k-k')$. $\omega_k$ is the frequency of travelling photon with a wave vector of $k$. The last term of the Hamiltonian describes the dipole interactions between travelling photons and each mode, which are linear in $\hat{p_k}$ and $\hat{p_k}^\dag$. $\lambda_m$ and $\lambda_a$ are individual coupling strengths of two modes with travelling photons. $\phi$ indicates the phase delay of the travelling photon from one mode to another, i.e., $\phi=kL$ where $L$ is the separation between two modes. 

\subsection{Derivation of the Langevin equations}

Following a standard procedure\cite{Clerk2010,Walls,Gardiner1985} and using the rotation-wave approximation, the equation of motion for the travelling photon $\hat{p_k}$ can be solved from Eq. (\ref{Eq1}):
\begin{equation}
\frac{d\hat{p_k}}{dt}=-\frac{i}{\hbar}[\hat{p_k},H]=-i\omega_k\hat{p_k}-i\lambda_me^{i\phi}\hat{m}-i\lambda_a\hat{a}
\label{Eq2}
\end{equation}
leading to
\begin{equation}
\hat{p_k}(t)=e^{-i\omega_k(t-t_0)}\hat{p_k}(t_0)-\int_{t_0}^ti[\lambda_me^{i\phi}\hat{m}+\lambda_a\hat{a}]e^{-i\omega_k(t-t')}dt'
\label{Eq3}
\end{equation}
where $\hat{p_k}(t_0)$ is $\hat{p_k}$ at the initial time of $t_0$ ($t_0<t$). Also from the Eq. (\ref{Eq1}), motion equations of the magnon mode and the other mode can be solved as:
\begin{eqnarray}
\frac{d\hat{m}}{dt}&=&-i\omega_m\hat{m}-\int i\lambda_me^{i\phi}\hat{p_k}dk \nonumber \\
\frac{d\hat{a}}{dt}&=&-i\omega_a\hat{a}-\int i\lambda_a\hat{p_k}dk
\label{Eq4}
\end{eqnarray}
Substituting Eq. (\ref{Eq3}) into Eq. (\ref{Eq4}), the derived quantum Lagevin equations of two modes are:
\begin{eqnarray}
\frac{d\hat{m}}{dt}&=&-i\omega_m\hat{m}-2\pi(\lambda_m^2\hat{m}+\lambda_m\lambda_ae^{i\phi}\hat{a})-i\sqrt{2\pi}\lambda_me^{i(\phi+\theta)}\hat{p}^{in} \nonumber \\
\frac{d\hat{a}}{dt}&=&-i\omega_a\hat{a}-2\pi(\lambda_a^2\hat{a}+\lambda_m\lambda_ae^{i\phi}\hat{m})-i\sqrt{2\pi}\lambda_a\hat{p}^{in}
\label{Eq5}
\end{eqnarray}
Here, we assume that $\lambda_m$ and $\lambda_a$ are independent of wave vector $k$. Therefore, damping terms in Eq. (\ref{Eq5}) depend only on two modes' operators evaluated at time $t$, and this arises from the first Markov approximation. In addition, considering that both two modes are two-side systems with mirror-symmetry boundaries, their damping rates induced by travelling photons have been doubled in Eq. (\ref{Eq5})\cite{Walls}. $\hat{p}^{in}$ is the input field which is defined as:
\begin{equation}
\hat{p}^{in}(t)=\frac{1}{\sqrt{2\pi}}\int e^{-i\omega_k(t-t_0)}\hat{p_k}(t_0)dk
\label{Eq6}
\end{equation}

In our assumptive system, two modes are separately placed in a waveguide. Input field ($\hat{p}^{in}$) arrives to each position at different time. Such a time delay has been considered by a travelling phase $\phi$. In addition, besides this delay, the phase shift of input field after passing through a resonant mode should also be considered. For instance, input field passes through the mode $\hat{a}(\hat{a}^\dag)$ first before arriving to the magnon mode. The phase of input field ($\hat{p}^{in}$) would be delayed by zero at $\omega_k\ll\omega_a$, $90^\circ$ at $\omega_k=\omega_a$ and $180^\circ$ at $\omega_k\gg\omega_a$. In Eq. (\ref{Eq5}), this resonant phase delay\cite{Resonantphase} is described by $\theta$. 

Radiative damping rates of the two modes are approximately defined as $\gamma\approx2\pi\lambda_m^2$ and $\kappa\approx2\pi\lambda_a^2$, respectively. Substituting them into Eq. (\ref{Eq5}), we get:
\begin{eqnarray}
\frac{d}{dt}
\left[\begin{array}{cc}
\hat{m}\\
\hat{a}\\
\end{array}\right]
 &=& -i\left[
  \begin{array}{cc}
  \omega_m-i(\alpha+\gamma) & -i\sqrt{\kappa\gamma}e^{i\phi} \\
  -i\sqrt{\kappa\gamma}e^{i\phi} & \omega_a-i(\beta+\kappa) \\
  \end{array}
  \right]
  \left[\begin{array}{cc}
\hat{m}\\
\hat{a}\\
\end{array}\right] \nonumber \\
&\quad&-i\left[\begin{array}{cc}
\sqrt{\gamma}e^{i(\phi+\theta)} \\
\sqrt{\kappa}\\
\end{array}\right]\hat{p}^{in}(t)
 \label{Eq7}
\end{eqnarray}
where $\alpha$ and $\beta$ respectively represent intrinsic damping rates of the two modes. Unlike the two aforementioned radiative damping rates, these two intrinsic damping rates have no effect on the interactions between two modes. They correspond to the inelastic scattering process of two resonant modes, and dissipate modes' energies into surroundings as heat. According to our theoretical model, coupling strength between these two separated modes is $-i\sqrt{\kappa\gamma}e^{i\phi}$. Obviously, when $\phi=n\pi$ ($n=0,1,2,3...$), it becomes an imaginary number, i.e., the coupling effect between two modes is purely dissipative.

\subsection{Transmission spectrum of system}

From Eq. (\ref{Eq2}), if we consider $t_1>t$, travelling photon $\hat{p_k}$ after interacting with two modes is:
\begin{equation}
\hat{p_k}(t)=e^{-i\omega_k(t-t_1)}\hat{p_k}(t_1)+\int_t^{t_1}i[\lambda_me^{i\phi}\hat{m}+\lambda_a\hat{a}]e^{-i\omega_k(t-t')}dt'
\label{Eq8}
\end{equation}
The output field is defined as:
\begin{equation}
\hat{p}^{out}(t)=\frac{1}{\sqrt{2\pi}}\int e^{-i\omega_k(t-t_1)}\hat{p_k}(t_1)dk
\label{Eq9}
\end{equation}
Carrying on the same procedures of Eq. (\ref{Eq4}) - (\ref{Eq7}), time-reversed Langevin equation of system can be derived, which describes the relation between two separated modes and the output field $\hat{p}^{out}$. Combining it with Eq. (\ref{Eq7}), we can obtain the input-output relation of system as:
\begin{equation}
\hat{p}^{out}(t)=\hat{p}^{in}(t)-i[\sqrt{\gamma}e^{i(\phi+\theta)}\hat{m}+\sqrt{\kappa}\hat{a}]
\label{Eq10}
\end{equation}

Using Eq. (\ref{Eq7}) and (\ref{Eq10}), transmission spectrum of system can be solved as:
\begin{eqnarray}
S_{21}&=&\frac{\hat{p}^{out}}{\hat{p}^{in}}=1-i(\begin{array}{cc}
  \sqrt{\gamma}e^{i(\phi+\theta)}, & \sqrt{\kappa}
\end{array})\centerdot \nonumber \\
&\quad&\left[
  \begin{array}{cc}
  \omega-\widetilde{\omega}_m & i\sqrt{\kappa\gamma}e^{i\phi} \\
  i\sqrt{\kappa\gamma}e^{i\phi} & \omega-\widetilde{\omega}_a \\
  \end{array}
  \right]^{-1}
  \left[\begin{array}{cc}
  \sqrt{\gamma}e^{i(\phi+\theta)} \\
  \sqrt{\kappa}
\end{array}\right]
\label{Eq11}
\end{eqnarray}
where $\widetilde{\omega}_m$ and $\widetilde{\omega}_a$ respectively represent two modes' complex frequencies, i.e., $\widetilde{\omega}_m=\omega_m-i(\alpha+\gamma)$ and $\widetilde{\omega}_a=\omega_a-i(\beta+\kappa)$. Because transmission spectrum can be directly measured from experiment, we choose it as the main study object in our following experiments. From Eq. (\ref{Eq11}), we note that transmission spectrum is a mixture of two different effects. Besides the indirect coupling effect with a strength of $-i\sqrt{\kappa\gamma}e^{i\phi}$, interference between two channels of input field $\hat{p}^{in}$ via $\sqrt{\gamma}e^{i(\phi+\theta)}$ and $\sqrt{\kappa}$ also exists.

\section{Experimental results}

\subsection{Characterization of the indirectly coupled cavity magnon system}

\begin{figure} [h!]
\begin{center}\
\epsfig{file=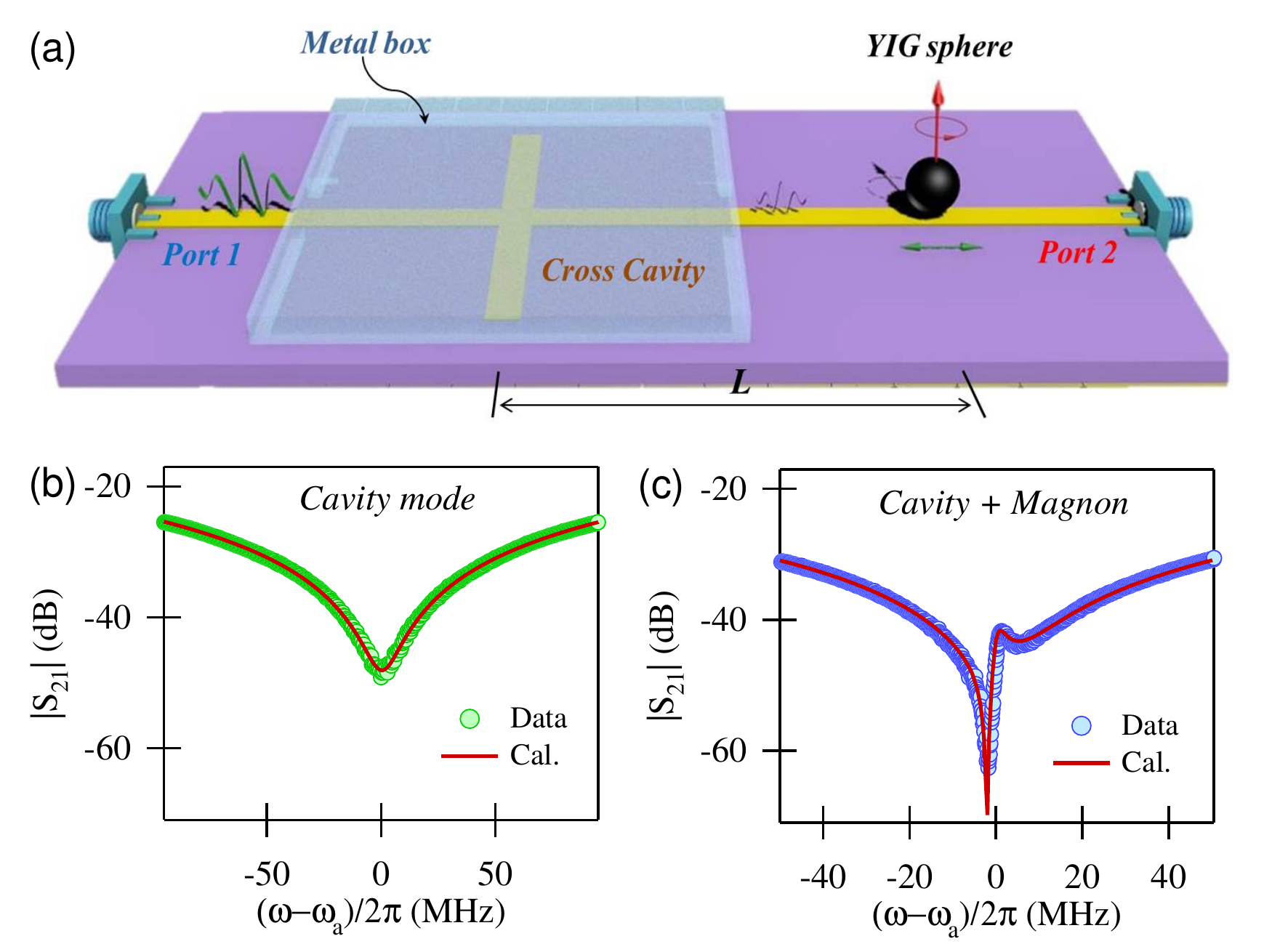,width=8.5 cm}
\caption{(a) Schematic diagram of the experimental set-up. The cross cavity is shielded by a metal box, so that it can only interact with the magnon mode via the transmission line. (b) Without the YIG sphere, the measured (green dots) and calculated (red solid line) transmission spectrum of system. (c) At the condition of $\omega_m=\omega_a$, the measured (blue dots) and calculated (red solid line) transmission spectrum.}
\label{fig1}
\end{center}
\end{figure}

Based on the theoretical model, we designed an experiment to test the indirect interaction between a magnon mode and a cavity photon mode in a planar waveguide. Figure \ref{fig1} (a) shows the experimental set-up, which contains three main parts: i) a cross cavity, which is shielded by a metal box; ii) a yttrium iron garnet sphere (YIG), whose position can be precisely controlled by a motor stage; iii) a 50 Ohm matched microstrip line, which is used to connect the cross cavity and the YIG sphere. 

The impedance matched microstrip line is fabricated on a 0.813 mm thick RO4003C substrate with a width of 1.67 mm. The cross cavity consists of two perpendicular arms, and is conductively connected to the microstrip line. The half-length of its vertical arm is 17.5 mm, which determines the mode frequencies of cavity. A 1 mm-diameter YIG sphere is placed on the top of the microstrip line and away from the cavity photon mode. A motor stage is used to move the YIG sphere along the microstrip line, and its variation range is from $L=25$ mm to $L=60$ mm. Because the cavity photon mode is shielded by the metal box, no direct mode overlap exists between it and the magnon mode. Considering both of these two modes are coupled to a mutual transmission line, any interaction arising between them can be attributed to the mediation of travelling photons. 

Without the YIG sphere, the transmission spectrum ($S_{21}$) of system is plotted in Fig. \ref{fig1} (b) as a function of $\Delta_a=\omega-\omega_a$. From it, radiative damping rate ($\kappa/2\pi$), intrinsic damping rate ($\beta/2\pi$) and mode frequency ($\omega_a/2\pi$) of the cross cavity are fitted as 1.77 GHz, 7 MHz and 2.775 GHz, respectively. The red solid line is the calculation result by using Eq. (\ref{Eq11}) with setting $\omega_m/2\pi=0$ GHz. After placing a YIG sphere at $L=41$ mm (Fig. \ref{fig1} (a)) and setting $\omega_a=\omega_m$, the measured transmission spectrum of system is plotted in Fig. \ref{fig1} (c). The two resonant dips in the spectrum are the result of the hybridization of the cavity photon mode and the magnon mode. Besides a small dip at $\Delta_a/2\pi=4$ MHz, an ultra-sharp dip\cite{Wang2019,Hsu2016} occurs at $\Delta_a/2\pi=-3$ MHz with an amplitude of nearly -70 dB. Still using Eq. (\ref{Eq11}), this spectrum can be well reproduced (red solid line in Fig. \ref{fig1} (c)) with setting $\gamma/2\pi=0.58$ kHz, $\alpha/2\pi=1.38$ MHz, $\phi=40.3^\circ$ and $\theta=90^\circ$. In this case, $\theta=90^\circ$ is because mode hybridization was measured at the condition of $\omega_a=\omega_m$. After passing through the cavity mode, the phase of input signal ($p^{in}$) would be shifted by a $90^\circ$.

\subsection{Indirect interaction between two separate modes at different separations}

\begin{figure} [h!]
\begin{center}\
\epsfig{file=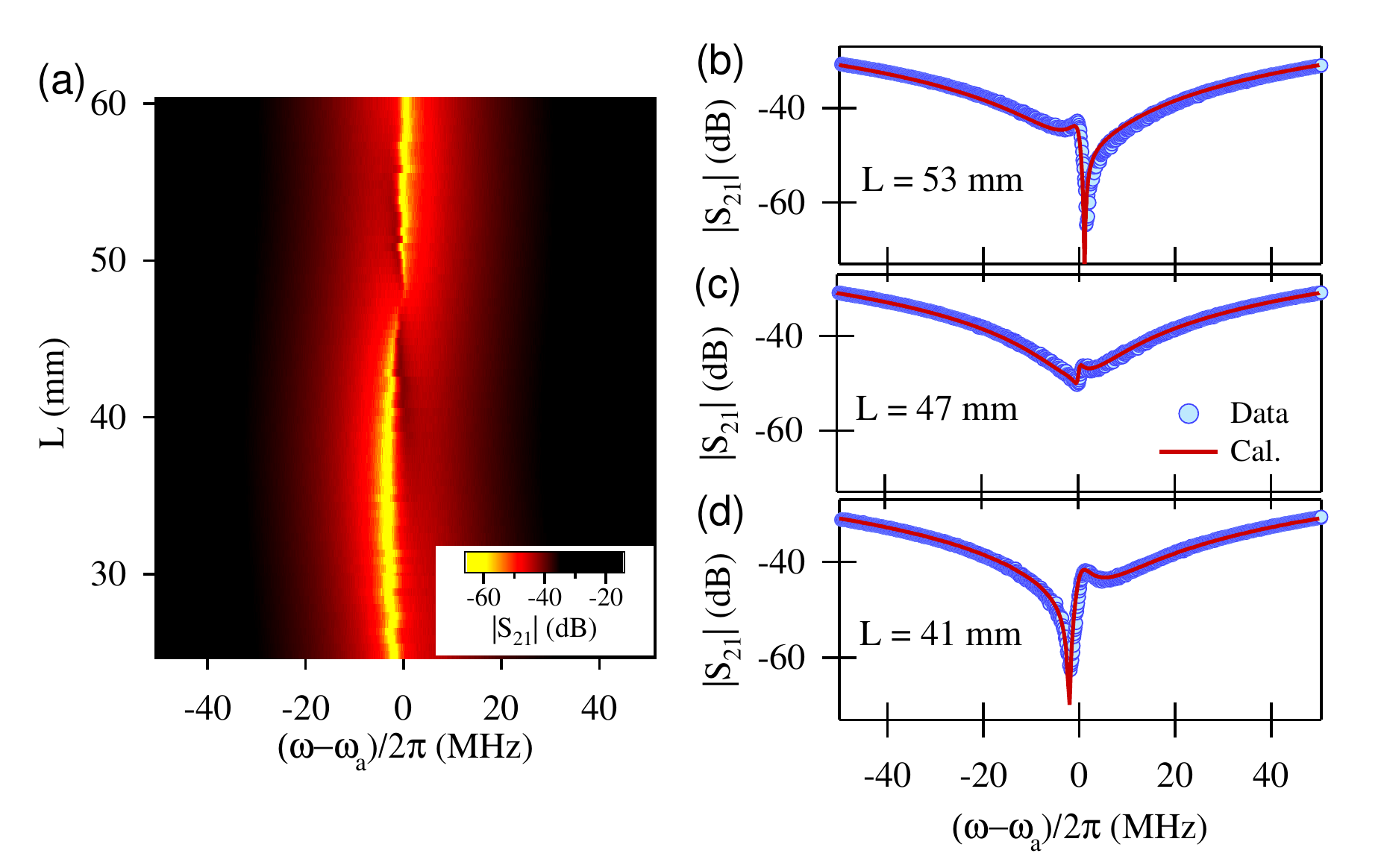,width=8.5 cm}
\caption{(a) Transmission of system ($|S_{21}|$) as a function of $\Delta_a$ and the separation between two modes $L$. (b)-(d) Three typical spectra measured at $L=53$ mm, $L=47$ mm and $L=41$ mm. Blue dots are experimental data and red solid lines are calculation results by using Eq. (\ref{Eq11}).}
\label{fig2}
\end{center}
\end{figure}

After characterizing the system, we set $\omega_m=\omega_a$, and measure transmission spectra of system at different positions of YIG sphere. Figure \ref{fig2} (a) shows the results, in which transmission are plotted as a function of $\Delta_a$ and the separation ($L$). The yellow color represents the aforementioned ultra-sharp resonant dip, which occurs from $-\Delta_a$ to $+\Delta_a$ with increasing the separation $L$, and nearly disappears at $L=47$ mm. Figure \ref{fig2} (b)-(d) are three typical spectra respectively measured at $L=53$ mm, $L=47$ mm and $L=41$ mm. By using Eq. (\ref{Eq11}), all these three spectra have been well described (red solid lines). 

\begin{figure} [h!]
\begin{center}\
\epsfig{file=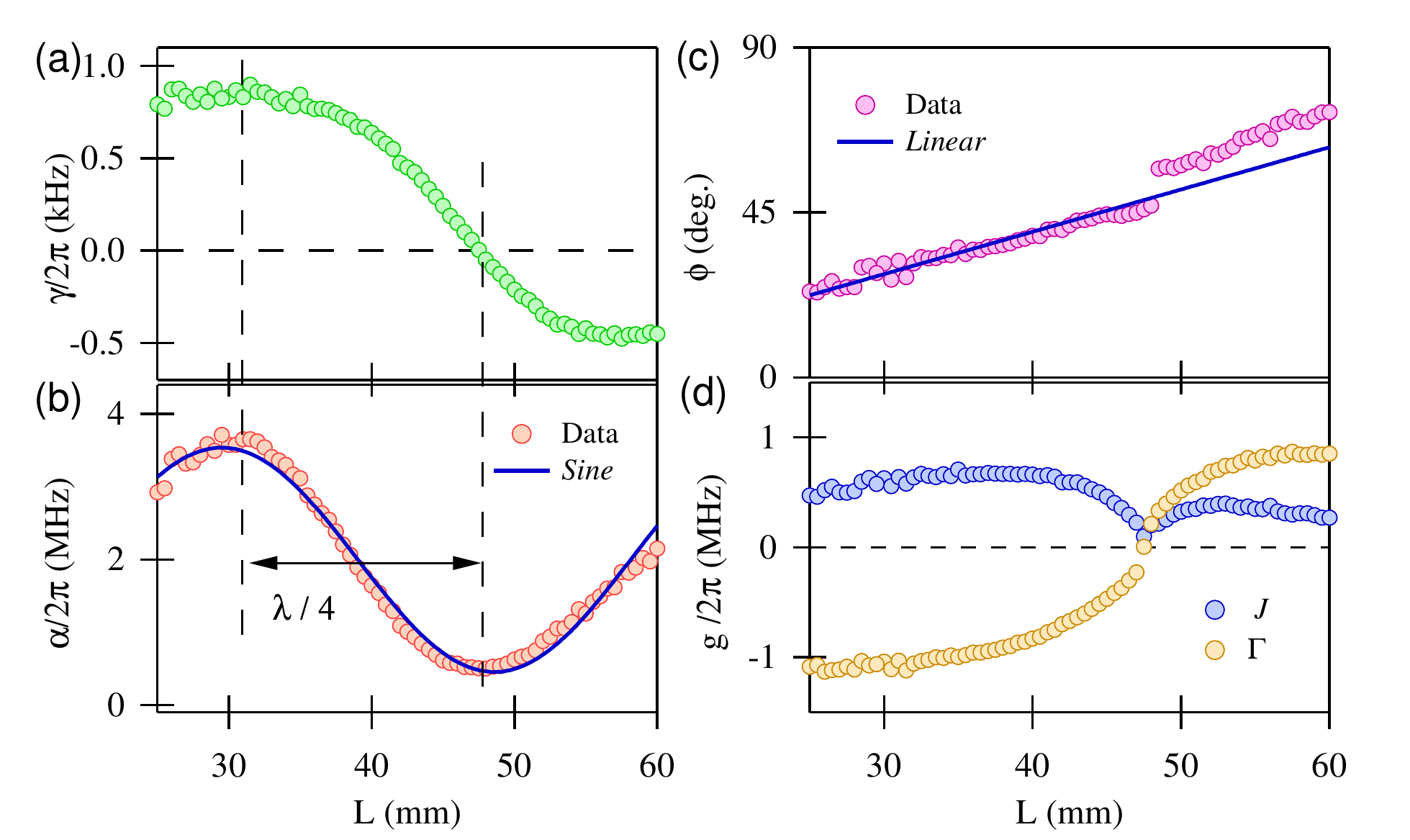,width=8.5 cm}
\caption{(a) and (b) Radiative damping rate ($\gamma$) and intrinsic damping rate ($\alpha$) of the magnon mode fitted from transmission spectra shown in Fig. \ref{fig2} (a). A sine function (blue solid line in (b)) is plotted to guide the eyes. The separation between the maximal and minimal values of two damping rates is indicated by two dashed lines, which equals to a quarter-wavelength of cavity photon mode. (c) Relative phase ($\phi$) as a linear function of the separation $L$. (d) Effective coherent coupling strength($J$) and dissipative coupling strength $\Gamma$ at different separations $L$.}
\label{fig3}
\end{center}
\end{figure}

Furthermore, we fit all spectra shown in Fig. \ref{fig2} (a), and plot the results in Fig. \ref{fig3} (a)-(c). The radiative damping rate $\gamma$ decreases from maximum to zero, and then becomes negative after $L=47$ mm (Fig. \ref{fig3} (a)). By contrast, the intrinsic damping rate of the magnon mode $\alpha$ (Fig. \ref{fig3} (b)) shows a sinusoidal dependence in separation $L$. It reaches the maximal value at $L=30$ mm, and then decease to minimal value at $L=47$ mm. As indicated by two dashed lines in Fig. \ref{fig3} (a) and (b), the separation between the maximal and minimal values of intrinsic damping rate is 17 mm, which approximately equals to the half-length of the vertical arm of cross cavity (17.5 mm), i.e., a quarter-wavelength of cavity photon mode\cite{Loo2013,Lalumiere2013}. Such a consistency between the spatial variation of damping rates and the wavelength of cavity photon mode indicates the influence on magnon system from the separate cavity photon mode.

In our system, the magnon mode interacts with travelling photons, which leads to relaxation, due to the emission of a photon at the resonant frequency, and a Lamb shift of mode frequency, due to the emission of virtual photons. Since both real and virtual photons can travel for a long distance in transmission line before being lost, they have a chance to be absorbed by the separate cavity photon mode when passing through it\cite{Loo2013,Lalumiere2013}. Likewise, the inverse process from cavity photon mode to the magnon mode is also true. Because of the exchange of real and virtual photons through a mutual transmission line, a non-trivial interaction between two separate modes arises. For the magnon mode, it emits photon to the transmission line, and, meanwhile, absorbs photon transmitted from the cavity photon mode. The residue of these two competitive processes results in the radiative damping of the magnon mode ($\gamma$). Depending on which process is dominant, the radiative damping of the magnon mode can be either positive or negative, as shown in Fig. \ref{fig3} (a). At $L=47$ mm, because two processes nearly cancel with each other, the radiative damping rate of the magnon mode vanishes. 

Additionally, for the spatial variation of intrinsic damping rate $\alpha$, we conjecture it may arise from the inelastic scattering process of the magnon mode. $\alpha$ contains two main parts: Gilbert damping and inelastic radiation. Gilbert damping of the magnon mode is determined by material's properties, which is independent of YIG's position. By contrast, the inelastic radiation of the magnon mode relates to the redistribution of waveguide electromagnetic field because of the existence of the cavity photon mode. When $\gamma$ reaches the maximal value at $L=30$ mm, energy dissipation of the magnon mode into both the travelling wave and the cavity photon mode reaches the maximal. Consequently, inelastic radiation of the magnon mode is enhanced, and $\alpha$ reaches its maximal value at this position. However, at $L=47$ mm, both two dissipation channels are shut down, because $\gamma$ nearly vanishes. In this condition, the magnon mode becomes decoupled with surroundings\cite{Loo2013}, so that inelastic radiation of the magnon mode nearly disappears. $\alpha$ falls back to the Gilbert damping rate around 0.5 MHz\cite{Yang2018}. 

Unlike two damping rates of the magnon mode, the travelling phase ($\phi$) exhibits a good linear relation with the separation $L$. Based on experimental data, $\phi$ at $L=0$ mm can be calculated as $-5^\circ(\approx0^\circ)$, which indicates coupling strength $-i\sqrt{\kappa\gamma}e^{i\phi}$ is an imaginary number at the intersection of the cross. In this condition, coupling effect between two modes is purely dissipative. This conclusion derived from our theoretical model is well consistent with previous works performed in the cross cavity\cite{Yang2019,Wang2019}. 

By using the extracted $\gamma$ and $\phi$ from curve fitting, effective coupling strength between two separated modes, i.e., $-i\sqrt{\kappa\gamma}e^{i\phi}=J+i\Gamma$, can be calculated (Fig. \ref{fig3} (d)). Here, $J$ and $\Gamma$ respectively represent the effectively coherent and dissipative coupling strengths between two modes. Note that the coherent coupling strength $J$ arises from the mediation of travelling wave, rather than the direct mode overlap of two modes in previous work\cite{Wang2019}. At the separation of $L=47$ mm, both $J$ and $\Gamma$ vanish, because $\gamma$ is zero. $J$ stays positive for all separations, but $\Gamma$ would change sign at $L=47$ mm. The switching sign of $\Gamma$ results in the observed ultra-sharp dip in transmission spectra changes from $-\Delta_a$ to $+\Delta_a$\cite{Wang2019}. 

\subsection{Mode hybridization at different bias magnetic field}

\begin{figure} [t!]
\begin{center}\
\epsfig{file=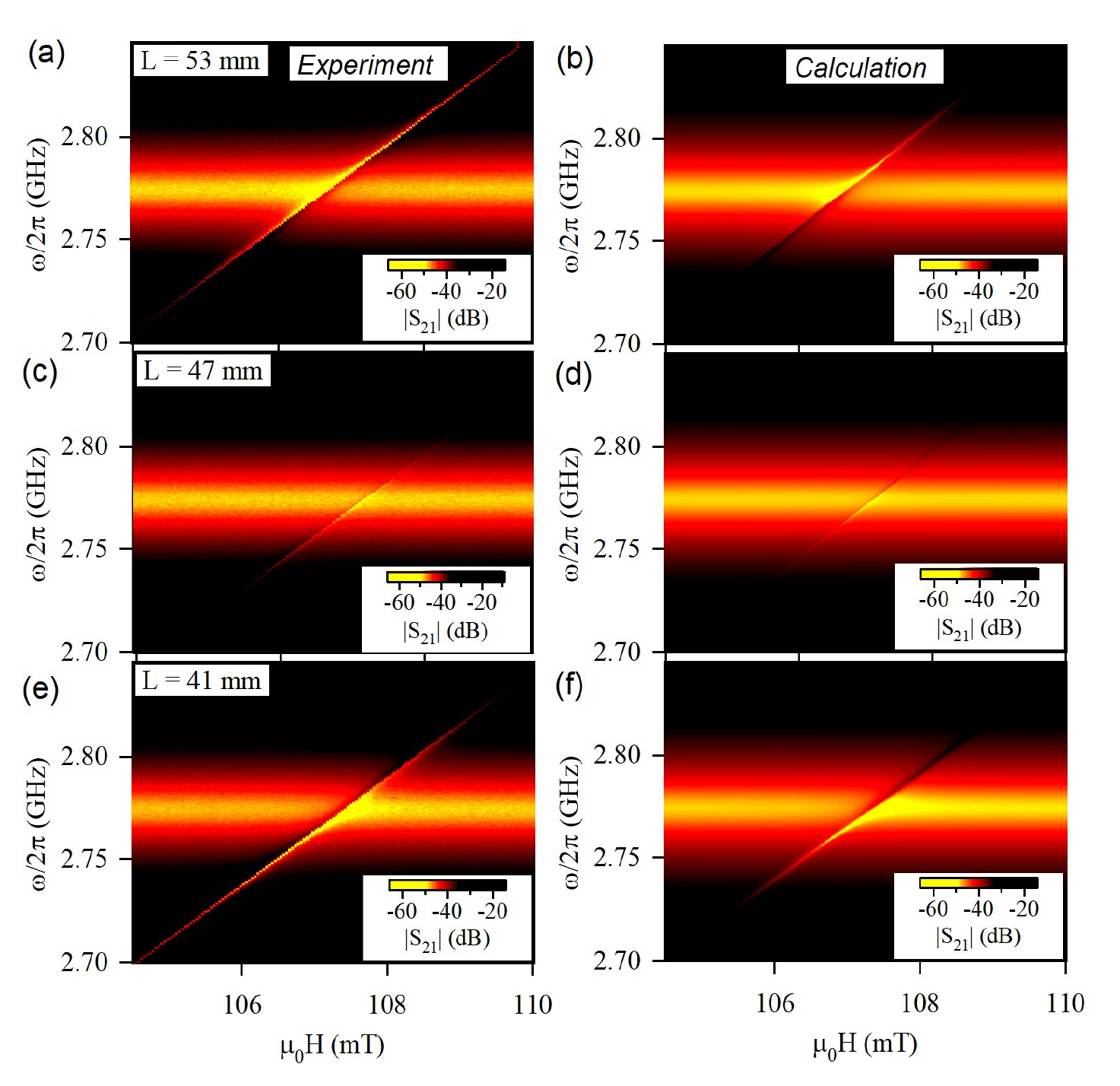,width=8.5 cm}
\caption{Typical transmission mappings at three positions. (a), (c) and (e) are experimental results respectively measured at $L=53$ mm, $L=47$ mm and $L=41$ mm. Correspondingly, (b), (d) and (f) are the calculation results.}
\label{fig4}
\end{center}
\end{figure}

To further understand the interaction behaviours between two separated modes, we have chosen three separations and measured transmission spectra of the system with sweeping the bias magnetic field (Fig. \ref{fig4} (a), (c) and (e)). From Fig. \ref{fig4} (a) and (e), a small gap occurs when frequencies of two modes matching with each other. In these two cases, coupling strengths $-i\sqrt{\kappa\gamma}e^{i\phi}$ are complex, so that mode hybridization of system exhibits as a mixture of level repulsion and level attraction. By contrast, at the separation of $L=47$ mm, because the radiative damping rate $\gamma$ vanishes, coupling strength of system decreases to almost zero. No obvious coupling gap can be observed in the transmission mapping (Fig. \ref{fig4} (c)). Figure \ref{fig4} (b), (d) and (f) are calculation results based on our theoretical model. The consistency between experimental data and theoretical calculation demonstrates that our theory built on the mediation of travelling photons grasps the main features of the indirect interaction between a cavity photon mode and the magnon mode. Away from the cavity mode frequency, difference between experimental data and calculations gradually emerges. The reason is that both $\phi$ and $\theta$ obtained at the condition of $\omega_a=\omega_m$ deviate from their actual values as $|\omega_a-\omega_m|$ increasing during calculations.

\section{Conclusion}

In this work, we studied the indirect interaction between a magnon mode and a cavity photon mode mediated by travelling photons. From a general Hamiltonian, we derived the effective coupling strength between two modes. It is determined by both values and relative phase of two modes' radiative damping rates, i.e., $-i\sqrt{\kappa\gamma}e^{i\phi}$. Therefore, at different separations, the coupling effect between them can be either coherent or dissipative. According to these theoretical predictions, we designed an experiment to test the interaction between a shield cavity photon mode and a magnon mode at different separations. Experimental observations confirm our theoretical predictions. Even without direct mode overlap, two modes can still couple with each other through their phase correlated dissipations into a mutual travelling photon bus. Our work demonstrates the indirect interaction mediated by travelling photons both in theory and experiment. The physics revealed in our work may help us develop the waveguide magnonics, and design new hybrid systems for quantum information processing.

\section*{Acknowledgements}
This work has been funded by NSERC Discovery Grants and NSERC Discovery Accelerator Supplements (C.-M. H.). J. W. Rao is supported by CSC scholarship. We would like to thank Yutong Zhao and Pengchao Xu for discussions, and also acknowledge CMC Microsystems for providing equipment that facilitated this research.


\begin{thebibliography}{199}
\bibitem{Resonantphase}
$\theta$ only affects the relative phase between two drive forces in Eq. (\ref{Eq5}), but has no influence on the effective coupling strength. In addition, because $\theta$ has no relation to the direction of input signal ($p^{in}$), it would not induce the non-reciprocity in our system.
\end{thebibliography}
\end{document}